\documentclass[11pt,a4paper]{article}
\usepackage[utf8]{inputenc}
\usepackage[english]{babel}
\usepackage{amsmath}
\usepackage{amsfonts}
\usepackage{amssymb}
\usepackage{lmodern}
\usepackage{geometry}
\usepackage{physics}
\usepackage{dsfont}
\usepackage{graphicx}
\usepackage[round]{natbib}
\usepackage{geometry}
\usepackage{wrapfig}
\usepackage{pgfplots}
\usepackage{tikz}
\usepackage{relsize}
\usepackage{titling}
\usepackage{hyperref}
\usetikzlibrary{decorations.pathmorphing}
\title{The multiple realizability of general relativity in quantum gravity}
\date{Forthcoming in \textit{Synthese}\thanks{This is a post-peer-review, pre-copyedit version of an article published in \textit{Synthese}. The final authenticated version is available online at: http://dx.doi.org/10.1007/s11229-019-02382-8}}
\author{Rasmus Jaksland\thanks{Department of Philosophy and Religious Studies, NTNU -- Norwegian University of Science and Technology, Trondheim, Norway. Email: rasmus.jaksland@ntnu.no}}
\begin{document}
\maketitle
\setlength{\parskip}{1mm plus0.5mm minus0.2mm}
\setlength{\parindent}{12pt}

\section*{Abstract}
Must a theory of quantum gravity have some truth to it if it can recover general relativity in some limit of the theory? This paper answers this question in the negative by indicating that general relativity is multiply realizable in quantum gravity. The argument is inspired by spacetime functionalism -- multiple realizability being a central tenet of functionalism -- and proceeds via three case studies: induced gravity, thermodynamic gravity, and entanglement gravity. In these, general relativity in the form of the Einstein field equations can be recovered from elements that are either manifestly multiply realizable or at least of the generic nature that is suggestive of functions.

If general relativity, as argued here, can inherit this multiple realizability, then a theory of quantum gravity can recover general relativity while being completely wrong about the posited microstructure. As a consequence, the recovery of general relativity cannot serve as the ultimate arbiter that decides which theory of quantum gravity that is worthy of pursuit, even though it is of course not irrelevant either qua quantum \textit{gravity}. Thus, the recovery of general relativity in string theory, for instance, does not guarantee that the stringy account of the world is on the right track; despite sentiments to the contrary among string theorists.

\subsection*{Keywords}
\textit{quantum gravity}, \textit{general relativity}, \textit{spacetime functionalism}, \textit{entanglement}, \textit{emergent gravity}, \textit{multiple realizability}, \textit{string theory}

\newpage

\section{Introduction}\label{Introduction}
Spacetime functionalism, as advocated by  \citet{knox_effective_2013,knox_physical_2017} with inspiration from \citet{brown_physical_2005}, is a prominent recent view of spacetime in general relativity and beyond. As its name suggests, spacetime functionalism identifies spacetime, not with a physical kind, but instead with a function: ``the metric field is spacetime because of what it does [...] and not by way of what it is" \citep[3]{knox_physical_2017}. The present paper finds inspiration in this functionalist conception of spacetime for the conjecture that general relativity as a whole is not linked to a particular physical kind, but rather takes the form of a role that might therefore be realized by multiple types of microstructure. In this light, the place of general relativity in the assessment of theories of quantum gravity is reconsidered. 
Based on three case studies, the paper finds reasons to expect that general relativity will prove to be multiply realizable among possible theories of quantum gravity; known or unknown. 
The paper argues on these grounds that we should not be too impressed by the recovery of general relativity in theories of quantum gravity.

This question proves important given the current state of quantum gravity research where any theory is yet to deliver testable empirical consequences. It is to be decided by non-empirical means which theory of quantum gravity that is worthy of pursuit and funding, and here stories about the successes of the theory are centre stage. One such story often alluded to in the string community is the natural occurrence within this theoretical framework of gravity and general relativity:\footnote{The reader is referred to \citet{dawid_underdetermination_2006,dawid_string_2013,dawid_theory_2013} for a more thorough analysis of non-empirical confirmation in the context of string theory. Consult \citet{dawid_conflicting_2009} for general epistemic assessment of the status of string theory.
}
\begin{quote}
String theorists sometimes say that string theory already made at least one successful \textit{prediction}: it predicted gravity! [...] There is a bit of jest in saying so -- after all, gravity is the oldest known force in nature. I believe, however, that there is a very substantial point to be made here. String theory is the quantum theory of a relativistic string. In no sense whatsoever is gravity put into string theory by hand. It is a complete surprise that gravity emerges in string theory. Indeed, none of the vibrations of the \textit{classical} relativistic string correspond to the particle of gravity. It is a truly remarkable fact that we find the particle of gravity among the \textit{quantum} vibrations of the relativistic string \citep[11]{zwiebach_first_2009}.
\end{quote}

\begin{quote}
The first general feature of string theory, and perhaps the most important, is that general relativity is naturally incorporated in the theory \citep[3]{becker_string_2007}.\footnote{See also \citet[145]{greene_elegant_1999} and \citet[5]{polchinski_string_1998}.}
\end{quote}
But how impressed should we actually be by the ``remarkable fact" that a theory of quantum gravity features the graviton and can ``incorporate" general relativity? These and other remarks by proponents of string theory can give the impression that the theory, as a result of these successes, is somehow guaranteed to have some truth to it. In one sense, this holds trivially: if general relativity is recovered in some limit of a theory of quantum gravity, then at least the theory is (apparently) empirically adequate in that regime. However, the confidence in parts of the string community -- implicit in the references above -- is indicative of the stronger sentiment that the successful recovery of general relativity proves string theory to be on the right track.\footnote{Indeed, `on the right track'-talk is a recurring theme among string theorists as documented by \citet{camilleri_role_2015}.} Obviously, this does not amount to the claim among its proponents that string theory in its current form is exactly true, but neither merely that it promises to be empirically adequate in some limit.\footnote{Thank you to an anonymous reviewer of \textit{Synthese} for pressing me on this point.} String theory provides a novel account of the fundamental constituents of reality by positing a microstructure of two-dimensional relativistic strings (and D-dimensional branes, possibly induced by the strings) from which general relativity is recovered as large scale behaviour. The conviction in parts of the string community seems to be that the successful recovery of general relativity promises that this theoretical framework will not prove to be a dead end in quantum gravity research; that some aspects of the stringy account of the world -- perhaps most likely those related to the recovery of general relativity -- will find their way to the theory of quantum gravity that will eventually be vindicated.\footnote{It is charitable to observe that string theory has other features to recommend it such as indications toward the recovery of the Standard Model via D-branes \citep{antoniadis_d-branes_2003}, providing the expected entropy for black holes \citep{strominger_microscopic_1996}, and renormalizability. Still, in most accounts the recovery of general relativity is presented as the prominent success of string theory.} In light of this, the question to be explored here is whether the recovery of general relativity in some limit of a prospective theory of quantum gravity  \textit{guarantees} that at least some elements of the theory -- in particular the microstructure it posits -- will find recognizable counterparts in the actual theory of quantum gravity of our universe. It will here be proposed that this question should be answered in the negative. 
Instead, it is contended that there are at least indications that the routes to general relativity in quantum gravity are rather generic. It is therefore to be expected that multiple theories of quantum gravity -- known or unknown -- with very different microstructure can recover general relativity. A theory can therefore recover general relativity while being completely wrong about the microstructure it posits; the recovery of general relativity is not a guarantee that string theory is on the right track. What is the true theory of quantum gravity will, according to this argument, not be decided on the grounds of which theory recovers general relativity.

The argument to this effect has functionalism as its point of departure and is inspired by the observation of \citet{lam_spacetime_2018} in the context of spacetime functionalism that ``it is seen in general as a virtue of functionalism that it allows -- indeed, accounts for -- multiple realizability" (5). 
Since functions are roles rather than physical kinds, they are realized not by what the realizer is, but by what it does. As such, functions are independent of the internal constitution of the realizer; they do not depend on the \textit{nature} of the microstructure that constitutes the function in a particular circumstance.\footnote{The functionalism assumed here is an ontological functionalism in the terminology of \citet{bealer_self-consciousness_1997}.} If something is a function, then this explains how multiple systems -- irrespective of how different their internal constitution might be -- can realize it; something just has to play the right functional role. It is particularly this aspect of functions that will inspire and inform the discussion of the recovery of general relativity in quantum gravity in general and string theory in particular. The idea is simply this: If sufficiently large parts of the complex described by general relativity display this insensitivity to the nature of the underlying microstructure, then general relativity might be multiply realizable in quantum gravity. 
Here the category `quantum gravity' broadly denotes the possibly vast class of known and unknown theories that -- in the terminology of \citet{linnemann_hints_2018} -- can be characterized as \textit{prospective}  microtheories: theories that posit a common microstructure from which both quantum mechanics and general relativity might potentially be recovered in such a way that the theory can account for their individual success.\footnote{Exactly what qualifies a theory as a prospective microtheory -- and thus a potential candidate of a theory that can reconcile general relativity and quantum mechanics -- shall not concern us here. The important point is simply that the class of theories of quantum gravity includes not only the actual theory of quantum gravity, but also microtheories of other possible worlds with varying similarity to our own.} Thus, the argument seeks to establish that multiple theories in this class of quantum gravity theories can be expected to recover general relativity while positing very different microstructures.
\footnote{Whether general relativity is multiply realizable \textit{within} any one theory of quantum gravity is not at issue here. See \citet{huggett_target_2017} for a discussion relating to this other theme in the context of string theory.} This ambition also explains the restriction of the class of quantum gravity theories to microtheories. While general relativity is trivially recovered in semi-classical gravity -- with or without higher order corrections -- and various effective field theories that include the Einstein-Hilbert action, such theories do not posit any microstructure and the recovery of general relativity is simply achieved by stipulation. Consequently, they cannot serve as evidence that general relativity can be recovered by multiple types of microstructure. For the purposes of this paper, they are not proper theories of quantum gravity.

The multiple realizability of general relativity directly bears on the significance of the recovery of general relativity by a theory of quantum gravity. If multiple theories of quantum gravity with different types of microstructure can recover general relativity, then a theory can evidently do so and still be completely misguided, i.e. it is possible that none of its elements, the posited microstructure in particular, will find their way to the true theory of quantum gravity of the actual world. The simplest route to this conclusion would be to show that several known prospective theories of quantum gravity do recover general relativity while positing very different microstructure. However, as argued in section \ref{Towards an argument} neither of these premises are clearly satisfied by the current field of prospective theories of quantum gravity. Instead, the strategy of section \ref{Three routes to the EFEs} is to show how general relativity -- or more precisely the Einstein field equations (EFEs) -- can be recovered from elements that are either manifestly multiply realizable or at least generic in nature in the varying circumstances of three case studies: induced gravity, thermodynamic gravity, and entanglement gravity. In all three cases, the recovery of the EFEs does not depend on the exact details of the microstructure realizing it: the recovery only depends on generic elements such as unspecified quantum fields, thermodynamic properties, and entanglement. Each approach indicates that the EFEs can inherit this insensitivity to the nature of the microstructure and with it, the possibility that systems with very different microstructure can realize them; having three different approaches to this effect only bolsters the conclusion.\footnote{Having three different, independent approaches leading to the same conclusion establishes its ``robustness" beyond what is achieved by each individual approach \citep{wimsatt_robustness_1981}. However, as observed by \citet{weisberg_robustness_2006}: ``Robustness analysis helps to identify robust theorems, but it does not confirm them" (742).} 
If this holds, i.e. if general relativity is multiply realizable in quantum gravity, then the ``remarkable fact that we find the particle of gravity among the \textit{quantum} vibrations of the relativistic string" and that ``general relativity is naturally incorporated in the theory" are no guarantee that the stringy account of the world is on the right track! More generally, the present paper argues that there are indications that theories of quantum gravity can recover general relativity without their posited microstructure being even approximately that of the actual world.

One might object that no one would actually claim that the recovery of general relativity implies even approximate truth, and that the conclusion is therefore trivial. I think the attitude in parts of the string community disproves this, but regardless, the present argument serves to inform a related important question also alluded to above: how to prioritize for instance funds and resources between different quantum gravity research programs. The argument of this paper indicates that we might have to reconsider the significance assigned to the recovery of general relativity in our preliminary assessment of theories of quantum gravity. The recovery of general relativity is one of the two primary desiderata for any theory of quantum gravity and as such, it does have some confirmatory significance in quantum gravity.\footnote{I would like to thank an anonymous reviewer of \textit{Synthese} for pressing me on this issue.} Rather, what this paper claims to inform is how we should update our credence in a prospective theory of quantum gravity upon the discovery that it recovers general relativity. If we use Bayes' theorem to update our credence in a theory of quantum gravity upon the discovery that it recovers general relativity, the prior probability of recovering general relativity in such theories, $P(GR)$, enters in the denominator: 
\begin{equation}
P(QG \vert GR) = \frac{P(GR \vert QG) P(QG)}{P(GR)}
\end{equation}
where $P(QG)$ is the prior credence in some theory of quantum gravity, $P(QG\vert GR)$ is the updated credence upon the discovery that it recovers general relativity, and $P(GR \vert QG)$ is the probability that it actually recovers general relativity (if the recovery of general relativity is a formal result, then this latter probability is arguably one). All are estimated within the context of the class of theories of quantum gravity. Arguably, this prior probability of recovering general relativity, $P(GR)$, is significantly larger if general relativity is multiply realizable, than if it is in a one-to-one relation with a specific (possibly unknown) microstructure. So if we update our credence in light of evidence based on a Bayesian inference, then our credence in a theory of quantum gravity that recovers general relativity, $P(QG \vert GR)$, is less than otherwise, if general relativity is multiply realizable, i.e. if the argument of this paper goes through. We should, in that case, not be too impressed by the recovery of general relativity in a theory of quantum gravity including string theory.\footnote{To question the significance of the recovery of general relativity in string theory is not in itself novel and has already been argued to be part of a larger debate where ``string theorists and their critics typically adopt different attitudes to the heuristic significance of solved, partially solved, and unsolved problems" \citep[54]{camilleri_role_2015}. However, when it comes to the recovery of general relativity in string theory, the dispute over its heuristic significance has only ensued as the debate whether this problem is solved or only partially solved by string theory (see footnote \ref{criticism}). In contrast, the present argument directly informs the heuristic significance of recovering general relativity in a theory of quantum gravity and contends that it should be moderated, if general relativity, as argued, is multiply realizable in quantum gravity.} 
As a result, 
the argument might adjust the relative weight of this and other factors -- UV-completion, universality, structural uniqueness, recovering quantum mechanics, to name some -- involved in prioritizing between different quantum gravity research programs\footnote{See the contributions to \citet{dardashti_why_2019} for some recent instalments in the debate over the role and relevance of such factors in theory assessment in fundamental physics.} in the context of pursuit.
\footnote{See \citet{cabrera_string_2018} for a discussion of the difference between the context of pursuit and the context of justification in relation to string theory.}
In summary, when this paper purports to indicate that general relativity is multiply realizable in quantum gravity, then this is not only meant to (possibly) update individuals' credence in current and future prospective theories of quantum gravity, but also to contribute to the discussion of the appropriate relative volume of research in the competing research programs.

The paper proceeds as follows: Section \ref{General relativity and string theory} reviews the recovery of general relativity in string theory and, based on this, suggests a sufficient condition for the recovery of general relativity -- that the EFEs can be derived in some limit -- which will be operational in the remainder of the paper. Section \ref{Multiple realizability} then details multiple realizability and how the string theory derivation does \textit{not} display this. Section \ref{Towards an argument} explores, but ultimately rejects, an argument that general relativity is multiply realizable based on its prospective recovery in different current theories of quantum gravity. Instead, section \ref{Three routes to the EFEs} gives an argument to the same effect based on the generic nature of the route to the EFEs in induced gravity (\ref{Induced gravity}), thermodynamic gravity (\ref{Thermodynamic gravity}), and entanglement gravity (\ref{Entanglement gravity}), respectively. The paper concludes that there are indications that general relativity is multiply realizable in quantum gravity.

\section{General relativity and string theory}\label{General relativity and string theory}
It is helpful to begin the discussion with a brief review of what string theory can actually do when it comes to recovering general relativity. This both serves to set the bar for what has so far been achieved by a theory of quantum gravity and suggests what to look for in other approaches that promise to deliver general relativity. For a rigorous treatment, see \citet{callan_strings_1985}. For a more philosophical exploration, see \citet{huggett_deriving_2015}.

First, a bit of set-up: consider a closed string, i.e. a string whose ends meet, on a Lorentzian manifold known as target space; due to consistency constraints, target space is 26-dimensional in bosonic string theory and 10-dimensional in superstring theory, but this shall not concern us here. Whereas a particle only extends in one dimension (time), a string extends in one spatial dimension in addition to the time dimension; strings are two-dimensional objects that span worldsheets instead of worldlines. The action for the relativistic point particle is its mass multiplied by the proper time along its worldline, and analogously, the string action is given in terms of the ``proper" area swept by the string and the mass per unit length of the string. For the relativistic point particle, the parametrization of the worldline entering the action holds no physical significance, and we therefore require that all physics must be invariant under reparametrization of this parameter. Similarly, the worldsheet is parametrized in terms of two coordinates which hold no physical significance either. String physics must therefore be invariant under a reparametrization of these worldsheet coordinates, but since the action also features the induced metric on the worldsheet that depends on these coordinates, the physics must in addition be Weyl invariant; invariant under change of local scale. Together, these are equivalent to conformal invariance, and any consistent string theory must therefore be conformally invariant. This entails that the field theory defined by the string action is a two-dimensional conformal field theory.

As usual, one can arrive at equations of motion by varying the action. Intuitively, the string equations of motion will admit various vibration modes of the string and indeed, in a particular gauge (a part of) the equations of motion take the form of a wave equation \citep[section 7.3]{zwiebach_first_2009}. Upon quantizing, these modes are quantum states that can be expressed in terms of creation and annihilation operators acting on a vacuum state. Among the first order excitation modes of the closed, relativistic, bosonic string are the graviton (a massless, symmetric, traceless tensor field with two Lorentz indices) and the dilaton (a massless scalar field). This is the ``remarkable fact" that gravity naturally occurs in string theory: ``We never put in a dynamical metric [...], yet somehow, the quantum states of the gravitational field have emerged!" \citep[292]{zwiebach_first_2009}. From the perspective of the relation between string theory and general relativity, the graviton mode is interesting since it perturbs the metric of the background manifold where the closed string lives. A curved background can thus be conceived as a coherent state of closed string graviton modes. The same goes for the dilaton, whose fluctuations, it turns out, also contribute to the metric perturbation as seen from the point of view of general relativity (known as the Einstein frame). 

The derivation of the EFEs begins with a string moving in such a background of a coherent state of closed strings. The derivation assumes a low-energy limit. Since energy -- in units $c=\hbar=1$ -- scales like inverse length, this is the limit where the energy is much smaller than the inverse of characteristic length scale of the theory, which in the case of string theory is the string length, $l_s$.  Thus, the low energy limit is equivalent to the limit where $l_s \rightarrow 0$. This entails that massive modes of the closed string can be ignored since they scale as $l_s^{-2}$ (in Minkowski background). In the limit $l_s \rightarrow 0$, the massive modes therefore decouple such that only the massless modes -- the graviton and the dilaton -- contribute.\footnote{Strictly speaking, the massless modes of the bosonic string also includes the Kalb-Ramond field, and this derivation of the EFEs can therefore be regarded as one where the Kalb-Ramond field is assumed to vanish. This, however, makes no difference for the present purposes.} In this limit, one can therefore use an effective action for the two-dimensional worldsheet that takes the form of an interacting two-dimensional field theory that couples to the background metric and the dilaton field. 

Using the path integral formalism, the generating functional associated with the string action takes the form of a path integral (functional integral) over worldsheet metrics of manifolds and embedding functions from the worldsheet to target space. The dilaton field can be divided into a fluctuating part and constant part equal to its vacuum expectation value that only depends on the worldsheet topology. Consequently, this constant part of the dilaton field can serve as an expansion parameter for an expansion of the generating functional in topologies; the leading order contribution coming from worldsheets that are topological spheres and first order correction from topological tori. A bit of imagining serves to suggest that worldsheets with spherical topology correspond to tree level and the tori to first loop order; something that is indeed confirmed by the identification of the string coupling, $g_s$, with the constant part of the dilaton field. Thus, we can take the limit $g_s \rightarrow 0$ in addition to the low energy limit $l_s \rightarrow 0$. Taking both limits, the generating functional only gets contributions from the metric field and the fluctuating part of the dilaton field and takes the form of a path integral over metrics of spherical worldsheet manifolds. 

However, even at tree level, observables (correlator functions) evaluated using this generating functional diverges. To regulate the divergences, a cut-off is introduced, and the fields will generally depend on this cut-off, and this threatens to break the conformal invariance of the world sheet. At tree level, conformal invariance is only restored if the derivative of the background fields with respect to the cut-off vanishes, i.e. if the background fields do not depend on the cut-off anywhere. From perturbation theory in string coupling and string length one can find expressions at tree level -- valid in the limit $g_s, l_s \rightarrow 0$ -- for the derivative with respect to the cut-off of the metric and dilaton fields. Using the constraint that these must vanish everywhere, one finds for a bosonic relativistic string:\footnote{See \citet{callan_string_1987} for higher order corrections.}
\begin{equation}\label{beta=0}
\begin{split}
0 & = \frac{D-26}{6 l_s^2} (\nabla \Phi)^2 - \frac{1}{2} \nabla_A \nabla^{A} \Phi \\
0 & = R_{A B} + 2 \nabla_A \nabla_B \Phi.
\end{split}
\end{equation}
where $\Phi$ is the dilaton field, $\nabla$ is a covariant derivative, $R_{A B}$ is the Ricci tensor, $D$ is the number of dimensions of target space and $A$ and $B$ are Lorentz tensor indices of target space. These two equations are constraints that the background fields must satisfy in order for the physics to be consistent with a relativistic string moving in the background. Interestingly, the constraints reduce to the vacuum EFEs, $R_{A B} = 0$, for constant $\Phi$ (since $\Phi$ only enters with derivatives). This is the first indication that these consistency constraints recover general relativity.

While the equations (\ref{beta=0}) were obtained as the renormalization group flow for the regulated world sheet action, they can also be regarded as equations of motion for the background fields through which the string propagates. We may therefore change the perspective and search instead for a target space action that has (\ref{beta=0}) as its equations of motion. The action arrived at has some similarities with the Einstein-Hilbert action, and upon combining the metric field, $g_{A B}$, with the fluctuating part of the dilaton field, $\tilde{\Phi}$,
\begin{equation}
\tilde{g}_{A B} = e^{-4 \tilde{\Phi}/(D-2)} g_{A B}
\end{equation}
one arrives at the Einstein-Hilbert action with a matter content comprising a kinetic term coming from the fluctuations of the dilaton field. Finally, the full EFEs are the equations of motion for the Einstein-Hilbert action: ``In short, conformal invariance of string theory as a worldsheet QFT entails general relativity for target space" \citep[1168]{huggett_deriving_2015}. In their reflection on the derivation, \citet{huggett_deriving_2015} argue that the ``result shows that graviton [...] coherent states must be related appropriately: by the EFEs" and observe that ``the field equations are simply low energy descriptions of the string itself" (1169). General relativity is recovered from consistency constraints as an approximation from first order perturbation theory in the string length and string coupling for the closed string.\footnote{\label{criticism}Critics argue that this derivation of the EFEs is still short of recovering general relativity, because it -- in perturbing around a fixed background -- is not properly background independent (e.g. \citet[185-186]{smolin_trouble_2006} and \citet{rovelli_critical_2013}). Some string theorists have simply denied this proposing a weaker notion of background independence, while others argue that this is merely an artefact of the background dependent perturbative string theory used for the derivation, but maintain that there are indications that the (unknown) non-perturbative formulation of string theory is background independent 
\citep{camilleri_role_2015}. We shall set this issue aside, but observe that the same worry applies to the three routes to the EFEs considered in section \ref{Three routes to the EFEs}. They may therefore also be background dependent and thus short of full derivations of general relativity.}

The derivation of the EFEs in string theory is the benchmark for the later exploration of other routes to general relativity in quantum gravity and it can thus serve as a guide for what it means when we say that a theory of quantum gravity recovers general relativity. 
In light of the string theory derivation, the following condition will be assumed to be \textit{sufficient} for recovering general relativity in quantum gravity:
\begin{quote}
\textbf{Definition:} A theory of quantum gravity is said to recover general relativity if the  theory describes a system comprising the whole universe with dynamics that in some coarse-grained limit are governed by the (semi-classical) EFEs.
\end{quote}
The qualification `semi-classical' is added to admit that the matter can couple to the metric of the spacetime via the expectation value of the energy-momentum operator.

The qualification that the system comprises the whole universe is included to deal with the worry that the uninterpreted EFEs merely relate components of two tensors to one another. We can describe many different waves -- water waves, electromagnetic waves, sound waves etc. -- using the wave equation, and so, one might object, could the dynamics of a system be governed by the EFEs, but only feature pseudo-gravity; something analogous to general relativity but qualitatively different such as analogue black holes \citep{weinfurtner_classical_2013}.\footnote{See \citet{dardashti_confirmation_2015} and \citet{crowther_what_2019} for a discussion of the significance of such analogue black holes. See \citet{barcelo_analogue_2011} for a general review of analogue gravity.} Such examples should not inform whether general relativity is multiply realizable since realizations, for instance in the form of analogue gravity, are instances where general relativity is also part of the best account of the system realizing this pseudo-gravity. To assess this objection, it is useful to adopt some terminology from the interpretation of dualities: the distinction between an internal and external point of view of a theory. Based on work by \citet{dieks_emergence_2015}, Sebastian \citet{de_haro_dualities_2017} defines ``[a]n external point of view, in which the meaning of the physical quantities is externally fixed" (116). This is contrasted with the internal point of view which obtains ``if the meaning of the symbols is not fixed beforehand" \citep[116]{de_haro_dualities_2017}. As de Haro observes, these labels are meant to track the available resources with which to interpret a theory: ``‘External’ suggests that there is a relevant environment to which the theory may be coupled; an external context that fixes the meaning of the symbols. ‘Internal’ suggests that such context is absent" \citep[116]{de_haro_dualities_2017}. If a theory is a theory of the whole world, de Haro argues, then only the internal point of view is possible. In that case, there is no external context with which to fix the meaning of the symbols entering the theory, i.e. to fix what physical quantity each symbol refers to. There is no external point of view since every point of view is encompassed by the theory. All there is to say about the system is given by the internal relations within the theory. Now, if two systems governed by the EFEs are claimed to be qualitatively different in such a way that one features gravity whereas the other only features pseudo-gravity, then this requires an external point of view: the possibility to assess that the symbols of the theory refer to different physical quantities between the two systems. Theories of quantum gravity, however, aspire to be theories of everything.\footnote{Though we will assume this throughout, more is, as observed by \citet{crowther_renormalizability_2017}, to be said about the relation between theories of quantum gravity, fundamentality, and the final theory of everything.} If EFEs govern the dynamics in some -- possibly coarse-grained -- limit of these theories, then the EFEs must inherit the original aspiration of the theory: to cover all elements of the universe, or at least those that are relevant in the limit covered by the EFEs. In quantum gravity, there is no external point of view, and it shall further be contended that this, consequently, is so for genuine recoveries of general relativity. Thus, general relativity is recovered whenever a theory of quantum gravity -- that by definition is a theory of the whole universe -- has some coarse-grained limit where the dynamics is governed by the EFEs.

In requiring the system to be \textit{governed by} the EFEs, the sufficient condition seeks to rule out cases where a model of the theory merely chances upon a Lorentzian manifold with matter content that happens to solve the EFEs. Suppose for instance that an empty universe model of a theory yields Minkowski spacetime -- a solution to the vacuum field equations. This is insufficient to recover general relativity, because the accordance with the EFEs could simply be an accident: there might another (possibly more generic) reason why the empty universe model yields Minkowski spacetime than that the metric curvature relates to the matter by the EFEs. To satisfy the proposed sufficient condition for recovering general relativity, Minkowski spacetime should relate to vacuum \textit{because} this is how curvature relates to matter in the EFEs (without cosmological constant). This qualification is already relevant in string theory, where the background target space is assumed to be some Lorentzian manifold and therefore one that, with the right matter content, might solve the EFEs. However, this would obviously not qualify as recovering general relativity, for the reasons above. It is also important to emphasize that assuming a Lorentzian background does not, by the same argument, amount to simply assuming general relativity. This qualification is important because all the three case studies of section \ref{Three routes to the EFEs} assume Lorentzian background. To satisfy the condition for recovering general relativity, it must be shown that each approach recovers the EFEs as the approximate dynamics in some limit, and thus that the EFEs are the equations of motion for the whole class of models that admit this approximation. In other words, while the case studies assume a Lorentzian manifold in their set-up, this does \textit{not} amount to assuming general relativity. Rather, the point of the case studies is exactly to argue that it follows from this and some additional generic assumption varying among the case studies that the relation between the curvature of the manifold's metric and the matter content is governed by the EFEs. As observed, this mirrors the derivation of the EFEs in string theory: the string theory derivation and the case studies are equal in this regard. The case studies are therefore telling of whether this type of derivation, and thus the recovery of general relativity as we now have it in string theory, can be expected in many prospective theories of quantum gravity.

\section{Multiple realizability}\label{Multiple realizability}
Like functionalism in general, multiple realizability is a most common theme in the philosophy of mind: ``The multiple realizability thesis about the mental is that a given psychological kind (like pain) can be realized by many distinct physical kinds"
\citep{bickle_multiple_2016}. The thesis, as we shall presently conceive of it, is that a mental state -- something at the psychological level of description -- has multiple realizations at the physical level of description.\footnote{Depending on one's view on reductionism in the context of philosophy of mind, the talk of levels of description might be regarded with suspicion. However, in the present context it seems uncontroversial that general relativity belongs to one length scale and quantum gravity to smaller length scale; at least if one does not promote this to a claim about ontological levels (see \citet{le_bihan_space_2018} for a discussion).} In this sense, multiple realizability goes against the one-to-one correspondence between mental states and brain states associated with identity theory in the philosophy of mind: ``The identity theory of mind holds that states and processes of the mind are identical to states and processes of the brain" \citep{smart_mind/brain_2017}. According to identity theory, all creatures -- or generally entities -- that realize some mental state must have a physical kind in common that this mental state can be identified with.\footnote{For identity theorists, this entails a identification between the mental state and its realizer, but this semantic component of the functionalism/identity theory debate shall not concern us here.}
 
In its contrast to functionalism, this \textit{identification} is the central thesis of identity theory: whereas functionalism argues that mental states are roles, identity theory argues that they a physical kinds. Similarly, one might imagine a similar debate about whether general relativity is to be identified with a role or a (complex of) physical kinds. With the present focus on multiple realizability, we shall set this semantic/metaphysical question aside, though we might observe (again) that multiple realizability sits well with functionalism. 
The contrast with identity theory here merely serves to raise the question whether there is a one-to-one or a one-to-many relation between the general relativistic level of description and that of quantum gravity.

In summary, the present paper has no ambition to say what general relativity \textit{is}, but only to assess whether we can expect to recover it in one or many theories of quantum gravity. What bears on this question is whether the recovery of general relativity relies on idiosyncrasies of a particular microstructure and its associated (type of) theory of quantum gravity, or whether general relativity can be recovered from generic elements expected to feature in multiple theories of quantum gravity. The derivation of the EFEs in the context of string theory can serve to illustrate this difference. From the perspective of multiple realizability, we should inquire about the elements entering this derivation and whether they themselves are multiply realizable or at least of a generic nature. We must, in other words, explore whether this derivation is insensitive to the exact nature of the microstructure at the level of description of quantum gravity. In this regard, the string theory derivation of the EFEs rules against multiple realizability. As shown above, string theory offers a route to the EFEs based on the Polyakov action in a curved background and consistency constraints for the worldsheet. But the Polyakov action is specifically the action for a relativistic object extended in time and one spatial dimension -- a string -- and the consistency constraints are imposed based on reasoning coming from reparametrization of the worldsheet swept by the string. If the string theory derivation were the only route to the EFEs, then it would seem that the EFEs were closely related to a particular sort of microstructure -- that of relativistic strings --  at the level of description of quantum gravity and thus in a one-to-one rather than one-to-many relation. As stated, the aim of this paper is to show that it might be otherwise: that we have reasons to expect there are many routes to the EFEs -- and thus the recovery of general relativity -- starting at different microstructures.

\section{Towards an argument}\label{Towards an argument}
As stated, the simplest route towards an argument for multiple realizability is to investigate whether different prospective theories of quantum gravity recover the EFEs in some limit. If they do, then this indicates that they can recover general relativity. If the theories in addition can be argued to exemplify different types of microstructure at the level of description of quantum gravity, then this would be evidence that general relativity is multiply realizable. At a glance, this approach has some promise. In causal set theory, a Lorentzian manifold with gravitational dynamics is conjectured to obtain as an approximation of discrete causal sets \citep{dowker_causal_2006}. In loop quantum gravity, something resembling the EFEs can be obtained by first finding a semi-classical approximation and then taking a classical limit \citep{wuthrich_raiders_2017}. In string theory, as shown above, the EFEs (with a particular field content) do obtain as equations of motion at first order perturbation theory in the string length and string coupling. Thus, an optimist might say that general relativity promises to occur in several prospective theories of quantum gravity. Furthermore, causal set theory, loop quantum gravity, and string theory could be argued to exemplify different types of microstructure since they, according to the quantum gravity taxonomy of \citet{huggett_emergent_2013}, are exemplars of three different types of theories: causal set theory is a non-metrical lattice theory, loop quantum gravity builds on superposed spin network states, and string theory is a quantum field theory that incorporates gravity.

This approach, however, faces immediate difficulties. First, the recovery of the EFEs is debated for all of the mentioned theories; even string theory.\footnote{For some recent instalments of this debate, see \citet{lam_dilemma_2013}, \citet{oriti_disappearance_2014}, \citet{carlip_challenges_2014}, and \citet{huggett_target_2017}.} While string theory has produced a rigorous derivation based on consistency requirements of the EFEs, this is in ten (or 26) rather than four dimensions. 
In covariant loop quantum gravity, the Regge action can be recovered from spinfoam amplitudes under certain additional, i.e. more than consistency, constraints. While the Regge action can be regarded as a type of discretized version of the Einstein-Hilbert action, there is no general mathematical proof that the Regge action converges to the Einstein-Hilbert action in the continuum limit (for more details, see \citet{han_einstein_2017} and references therein). Though promising, this is still short of a rigorous derivation of the EFEs in loop quantum gravity. 
In causal set theory, recovering the EFEs in some limit is still only a hope, and so is a formulation of its quantum dynamics. Also the independence of these prospective theories of quantum gravity has been cast into doubt. This questions the argument that these theories of quantum gravity really exemplify different types of microstructure. There are indications of a deep relation between string theory and loop quantum gravity \citep{jackson_conformal_2015,mertens_solving_2017}, and one can only speculate whether similar relations might obtain between these and causal set theory. The approach towards multiple realizability from the realization in different theories of quantum gravity relies on finding theories different enough to at least render it very unlikely that they share a common microstructure. The more general tenet of this second worry is therefore that the (conjectured) realization of general relativity in some prominent theories of quantum gravity might merely be an artifact of the type of theories we can imagine and find it worthwhile to pursue.

Again, an analogy to the case of consciousness can help exemplify this issue. Suppose you and your friend for some time have had the discussion whether consciousness is multiply realizable. You know that your friend is soon to attend the huge Intergalactic Gathering for All Creatures (IGAC) and you challenge her therefore to bring home different (apparently) conscious creatures to prove her point that consciousness is multiply realizable. You wait for her return in great anticipation, and with increasing excitement upon her message that she is bringing home three creatures from the IGAC. You imagine they could be weird slimy things, semi-transparent wisps, or green men with antennas. Big is your disappointment, therefore, when she arrives accompanied by a dog, a chimpanzee and your mutual friend Jack who also attended the IGAC. This of course does nothing to persuade you that consciousness is multiply realizable. First of all, that these were a selection of creatures from the IGAC makes no difference once you realize that all of them are from Earth. It seems likely that the dog, the chimpanzee, and Jack are not different enough to indicate multiple realizability and furthermore, while you are rather confident that Jack is conscious, you are less sure about the chimpanzee, and even less so about the dog. Presumably, these are very particular creatures among all the many different creatures attending the IGAC. They were probably chosen by your friend because they were easily recognized as possibly conscious creatures. Many of the other lifeforms present at the IGAC must have been very alien perhaps to the degree that it was difficult to assess whether they were alive at all, let alone conscious. So despite the initial disappointment, you can understand why your friend brought such mundane creatures home from the IGAC: these were the sort of creatures she was looking for. But precisely for this reason, they must be viewed with suspicion if used as a defence of the multiple realizability of consciousness. Being the sort of creatures she was looking for makes it all too likely that they do share a common physical/chemical kind that then is the realizer of consciousness. 

This story should serve as a warning when it comes to multiple realizability of general relativity in quantum gravity. For all we know, the class of prospective theories of quantum gravity is vast, and it seems overconfident to presume that causal set theory, loop quantum gravity, and string theory are the slime, wisp, and Martian of quantum gravity rather than the dog, the chimpanzee, and Jack. Just because these are theories out of the vast class of quantum gravity theories, they may well be too similar to tell us anything about multiple realizability; because these are the kind of theories we can imagine to look for. With this possibility, it is unjustified to rely too heavily on the particulars of these most prominent theories in the argument for the multiple realizability of general relativity. Instead, it seems more justifiable to look at general aspects of quantum gravity and whether these can be indicative of an answer to the inquiry. As such, the account finds inspiration in the arguments for emergent gravity reviewed by \citet{linnemann_hints_2018} that ``do not assume any specific QG model, but rather consist of pointing out features of gravity which are (at least typically) characteristic of systems with underlying microstructure" (2). As might be expected, this approach provides more implicit answers to the question of multiple realizability, but arguably it does so on epistemologically safer ground. 

\section{Three routes to the EFEs}\label{Three routes to the EFEs}
The argument involves three case studies. The first appeals to Sakharov's (\citeyear{sakharov_vacuum_1967}) induced gravity: that the Einstein-Hilbert action, and thus the EFEs, automatically emerges at first loop order of quantum field theories on Lorentzian manifolds \citep{visser_sakharovs_2002}. This is used to argue that the EFEs are generic to theories of quantum gravity. The second builds on the discovery by Ted \citet{jacobson_thermodynamics_1995} of a relation between thermodynamics and the EFEs.\footnote{See also \citet{padmanabhan_thermodynamical_2010} and \citet{verlinde_origin_2011}.}
Due to the versatility of thermodynamics, it will be argued that this relation establishes the possibility that general relativity is multiply realizable. Inspired by Jacobson's pioneering work, indications of a close relation between entanglement constraints and the EFEs in quantum gravity have appeared.\footnote{See for instance \citet{ryu_holographic_2006}, \citet{van_raamsdonk_building_2010}, \citet{maldacena_cool_2013},
\citet{chirco_spacetime_2014}, \citet{lashkari_gravitational_2014}, \citet{jacobson_entanglement_2016}, \citet{han_loop_2017}, \citet{baytas_gluing_2018}, \citet{cao_bulk_2018}, and \citet{chirco_group_2018}.} Arguing that entanglement is a generic and necessary component in any theory of quantum gravity, the relation between entanglement and gravity serves as yet another route to the multiple realizability of general relativity in quantum gravity.

Notice that the argument is \textit{not} simply an inference to multiple realizability from the recovery of general relativity in these three accounts in the fashion of the argument above based on causal set theory, loop quantum gravity, and string theory.\footnote{It could not be since none of the three approaches are theories of quantum gravity, but merely intermediate structures from which the EFEs can be derived.} Instead, the argument is meant to show that thermodynamic gravity -- and similarly for induced gravity and entanglement gravity -- is insensitive to the nature of the degrees of freedom realizing it and that general relativity can therefore be recovered as a thermodynamic phenomenon in systems with very different microstructure. Because of the possible thermodynamic nature of general relativity and the multiple realizability of thermodynamics, general relativity is multiply realizable. In this fashion, each account on its own indicates that general relativity is multiply realizable due to its origin in each account in multiply realizable elements or at least elements insensitive to the exact nature of the microstructure: first order quantum corrections in induced gravity, thermodynamics, and entanglement, respectively.

As mentioned in section \ref{General relativity and string theory}, all three accounts play out on a Lorentzian manifold, and none of them can as a consequence testify whether Lorentzian manifolds are multiply realizable. Indeed, both induced gravity and thermodynamic gravity are grouped by \citet{carlip_challenges_2014} as ``models in which spacetime, the
metric, and diffeomorphism invariance are present, but the
dynamics of gravity is emergent" (201); and the same holds for entanglement gravity.
As a consequence, they arguably only suffice to support that gravitational dynamics is insensitive to the microstructure realizing it and therefore possibly multiply realizable. For all said here, the assumed Lorentzian manifolds might be in a one-to-one correspondence with a particular microstructure at the level of description of quantum gravity. We will set this worry aside since it applies to the string theory derivation as well as the three case studies below. However, in the broader context of quantum gravity, of course, assuming a Lorentzian manifold is by no means innocent: Recovering a Lorentzian manifold is a most prominent problem in loop quantum gravity \citep{wuthrich_raiders_2017} and causal set theory \citep{dowker_causal_2006} and often equated with the issue of how spacetime emerges in quantum gravity (see for instance \citet{huggett_emergent_2013} and \citet{oriti_disappearance_2014}). Again, neither of these accounts hold any implications for this question about the emergence of a Lorentzian manifold; though recent research suggests a relation between spacetime and entanglement coming out of entanglement gravity \citep{van_raamsdonk_building_2010,van_raamsdonk_patchwork_2011,bianchi_architecture_2014,nomura_spacetime_2016}.

\subsection{Induced gravity}\label{Induced gravity}
Induced gravity is a form of emergent gravity\footnote{For a survey of emergent gravity, see \citet{linnemann_hints_2018}.} that pursues the idea that gravitational dynamics originate in or are \textit{induced} by quantum corrections from the matter fields. The programme originates in the discovery by \citet{sakharov_vacuum_1967} that semi-classical gravity in the form of the Einstein-Hilbert action necessarily occurs as part of the effective action in any quantum field theory on a Lorentzian manifold at first loop order, i.e. as first order quantum corrections to the classical equations of motion. In a few more words, the argument assumes a dynamical Lorentzian manifold with a generic QFT defined on it; nothing is assumed about the particulars of the field content of the QFT. The dynamics of the manifold is not specified other than that it couples to the matter content of the QFT. Based on this, one finds that the first order quantum corrections to the effective action -- contributions from one-loop Feynman diagrams -- necessarily include terms of the same form as the Einstein-Hilbert action from which the EFEs can be derived as equations of motion.\footnote{In the reconstruction by \citet{visser_sakharovs_2002}, the metric, $g_{ab}$, is expanded as the sum of a background, $g^{0}_{ab}$, and a perturbation, $h_{ab}(x)$ (nothing is assumed about the size of the perturbation). For the purposes of a Feynman diagram picture, the Fourier transform of the metric perturbation, $h_{ab}(k)$, takes the form of external gravitons. Sakharov's result is then that one-loop diagrams with an arbitrary number of external gravitons give rise to terms in the effective action of the form:
\[ \int d^4 x \sqrt{-g} \left[ c_1 + c_2 R(g) \right] \]
where $R(g)$ is the Ricci tensor, and $c_1,c_2$ are dimensionless constants that depend on the particular particle species. Compare this to the Einstein-Hilbert action:
\[
\int d^4 x \sqrt{-g} \left[ - \Lambda - \frac{R(g)}{16 \pi G_N} \right]
\]
where $ \Lambda $ is the cosmological constant. See \citet[5]{visser_sakharovs_2002} for further details on the connection between these two actions.} Thus, even if we do not have any gravitational dynamics in the form of EFEs or similar at tree level, i.e. at zero loops, it will occur at one-loop order in any QFT on a dynamical Lorentzian manifold. In a limit of the theory with one-loop dominance,\footnote{See \citet{visser_sakharovs_2002} for a survey of different routes to such a limit.} the effective action has the EFEs as equations of motion and the theory therefore recovers general relativity according to the condition of section \ref{General relativity and string theory}. \citet{visser_sakharovs_2002} summarizes the implications in the following manner: 
\begin{quote}
If we consider any candidate
theory for quantum gravity (brane models, quantum geometry, lattice quantum
gravity) then Sakharov's scenario tells us that trying to derive the inverse square
law from first principles is not the difficult step. Deriving the inverse square
law, and all of Einstein gravity is in fact automatic once you have demonstrated
the existence of Lorentzian manifolds (990). 
\end{quote}
The EFEs are generic equations of motion for QFTs on dynamical Lorentzian manifolds. Very little must be assumed about the details of the theory, for instance about the field content. Though it is confined to the QFT framework and dependent on the assumption of a Lorentzian manifold, Sakharov's scenario indicates, as signified by Visser, that the EFEs are generic in such theories. In other words, recovering general relativity by this route does not depend on the details of the underlying microstructure. This is the first indication that general relativity is multiply realizable. The point is not that induced gravity is the most promising approach to quantum gravity,\footnote{Indeed, the Weinberg-Witten (\citeyear{weinberg_limits_1980}) theorem suggests that induced gravity scenarios faces serious difficulties. See \citet{sindoni_emergent_2012} for a discussion.} but rather that induced gravity signifies how generic the EFEs are. 

\subsection{Thermodynamic gravity}\label{Thermodynamic gravity}
Jacobson's (\citeyear{jacobson_thermodynamics_1995}) thermodynamic gravity\footnote{There are many approaches going by the name of thermodynamic gravity all building to some extent on Jacobson's insight. See \citet{padmanabhan_thermodynamical_2010} for a review.}
is similar to induced gravity in that it derives the EFEs from few general ingredients and without adding gravitational dynamics to the mix by hand. Thermodynamic gravity relies on two basic assumptions: a proportionality between entropy, $S$, and area and a Clausius-like relation\footnote{This is an equilibrium thermodynamic relation.} $\delta Q = T d S$ connecting heat, $Q$, and temperature, $T$, to entropy. In Jacobson's thermodynamic gravity, the heat flow, $\delta Q$, is qualified as the energy flow over the causal horizon as observed by a boosted observer, known as a Rindler horizon.\footnote{The most well known example of a causal horizon is the event horizon of a black hole. The type of causal horizons considered here are those observed by a constantly accelerating observer as known from the \citet{unruh_notes_1976} effect.} It is therefore given as an integral over the energy-momentum tensor contracted with the local boost Killing vector field generating the Rindler horizon, and an area element of the Rindler horizon. Since causal horizons hide information, they can be associated with an entropy.\footnote{Again this is well known from black hole thermodynamics \citep{bekenstein_black_1973}.} The change in entropy, $d S$, associated with a piece of horizon is, by the assumed relation between entropy and area, proportional to the variation of the area of that piece of horizon. The change of area for a Rindler horizon can in turn be expressed as an integral of the Ricci tensor, the local boost Killing vector field generating the Rindler horizon and an area element of the horizon. Then, by inserting the integrals into the assumed equilibrium relation, $\delta Q = T d S$ and imposing energy-momentum conservation, the EFEs can be derived. \citet{jacobson_thermodynamics_1995} concludes: ``Viewed in this way, the Einstein equation is an equation of state" (1260). 

Evidently, a number of assumptions go into this derivation, and their truth is not directly the concern. Rather, what is of interest is that they are compatible with different types of microstructure, and thus sufficiently independent of the nature of the underlying microstructure. Heat, temperature, and entropy are all thermodynamic quantities related to the mean collective behaviour of underlying degrees of freedom.\footnote{The rigorous foundation of the relation between the macroscopic thermodynamic level of description and the microstructure on which it supervenes remains elusive \citep{sklar_physics_1993,ridderbos_coarse-graining_2002}. Here it suffices to observe that thermodynamics somehow does work to make reliable predictions.} They are macroscopic quantities related to macroscopic phenomena, visible only at a sufficiently coarse-grained level. In thermodynamic gravity, this insensitivity to the nature of the microstructure is inherited by the EFEs: ``gravity is viewed as robust with respect to the underlying microstructure, in the sense that it only
depends on coarse-grained notions such as entropy and temperature" \citep[4]{linnemann_hints_2018}. Indeed, the central characteristic of equilibrium thermodynamics is that it is sufficiently insensitive to the individual microscopic degrees of freedom; for instance the momentum of individual particles: ``thermodynamics was developed and used effectively decades before we understood the molecular structure of matter or its statistical mechanics" \citep[31]{padmanabhan_thermodynamical_2010}. This is signified by temperature -- one of the central elements is Jacobson's derivation of the EFEs -- being often portrayed as an exemplar of a multiply realizable function (e.g. \citet{kim_mind_1998,bickle_multiple_2016}). Thermodynamics is multiply realizable.

An equilibrium thermodynamic description is only available if the macroscopic behaviour of a system is sufficiently robust with respect to the individual changes among the microscopic degrees of freedom; a thermodynamic equilibrium is achieved when ``all reasonable macroscopic observables have
steady values" \citep[432]{pitowsky_definition_2006}.\footnote{This and many other aspects in the foundations of equilibrium thermodynamics are disputed. These will not be discussed further here. For a review, see \citet{frigg_field_2008}.} Jacobson is well aware of this restriction to thermodynamic gravity: ``It is born in the thermodynamic limit as a relation between thermodynamic variables, and its validity is seen to depend on the existence of local equilibrium conditions" \citep{jacobson_thermodynamics_1995}. However, from the present perspective, this signifies that this approach to obtain the EFEs is genuinely thermodynamic. So long as the microscopic degrees of freedom have macroscopic equilibrium states\footnote{See \citet{sindoni_horizon_2013} for further details on how such an equilibration might obtain.} that admit a conception of heat in the form of energy transfer over causal horizons and entropy as causally hidden information that is related to the horizon area, then the EFEs can be obtained as thermodynamic equations of state and general relativity can as a consequence be recovered as a thermodynamic phenomenon. The existence of causal horizons in the form of Rindler horizons is simply guaranteed by the assumption in the approach of a Lorentzian manifold and the possible existence of boosted observers. The accelerated observer perceives a causal horizon and it is therefore mandatory to integrate out the degrees of freedom causally hidden by the horizon which associates an entropy to the horizon. When matter crosses the horizon, then the entropy is changed according to the amount of energy transferred and the temperature of the reservoir which equates the temperature of the horizon: $\delta Q = T d S$.\footnote{See \citet[section 4]{padmanabhan_thermodynamical_2010} for further details. In essence, Jacobson's thermodynamic gravity derives from the thermodynamic nature of causal horizons. While horizon thermodynamics can be derived from the assumption that the EFEs are equations of state (e.g. \citet{hansen_horizon_2017}), horizon thermodynamics can also be defended on independent grounds and then shown to recover general relativity \citep{padmanabhan_gravity_2005}.} The proportionality between entropy and horizon area is perhaps the assumption of Jacobson's thermodynamic gravity that looks the most restrictive and consequently the one that might hinder the EFEs from inheriting the versatility of thermodynamics. However, a proportionality between entropy and horizon area is known from black hole thermodynamics \citep{bekenstein_black_1973}. And the close relation between black hole and horizon thermodynamics suggests that such a proportionality should obtain for all causal horizons \citep[section 7]{padmanabhan_gravity_2005}.\footnote{This argument by analogy to the proportionality between entropy and horizon area is at risk of being circular since the derivation of black hole thermodynamics assumes gravity. We shall here ignore this concern for two reasons: First, since there are routes to thermodynamic gravity that do not assume a proportionality between entropy and horizon area (e.g. \citet{padmanabhan_equipartition_2010}) and second, that the proportionality can be defended on independent grounds if the entropy is identified as entanglement entropy (more on this below). Observe that a similar concern arises if the proportionality is defended with reference to the holographic principle as formulated by \citet{t_hooft_dimensional_1994}, \citet{susskind_world_1995}, and  \citet{bousso_holographic_2002} since this principle also ultimately relies on black hole thermodynamics.} 
Black hole thermodynamics, in turn, is argued by \citet{wallace_case_2018} to be genuinely thermodynamic,\footnote{This view is disputed by \citet{dougherty_black_2016}.} and in so far as the proportionality between entropy and horizon area inherits this characteristic, it should also be sufficiently generic for thermodynamic gravity to inherit the multiple realizability of its thermodynamic origin:
\begin{quote}
Just as the bulk properties of solids can be described without reference to the underlying atomic structure, much of classical and semi classical gravity (including the entropy of black holes) will be independent of the underlying description of the microscopic degrees of freedom \citep[114]{padmanabhan_gravity_2005}. 
\end{quote}
No special assumption has to be made about the nature of the microstructure that realizes thermodynamic gravity; it relies solely on very generic features. Thermodynamic gravity is as generic as other forms of thermodynamics, and presuming that many types of quantum gravity microstructure can realize thermodynamics, this shows a way in which general relativity can be multiply realizable.

In summary, this possible thermodynamic origin of general relativity corroborates the view that general relativity is multiply realizable in quantum gravity. The derivation shows how the EFEs can inherit the multiple realizability of temperature and other thermodynamic properties. Thermodynamics can be realized on systems with very different types of microstructure; presumably including those posited by theories of quantum gravity. Thus, since general relativity, by the argument above, can be realized as a thermodynamic effect, and thermodynamics is likely to be multiply realizable in quantum gravity, then so is general relativity. The recovery of general relativity in a theory of quantum gravity does not entail that the theory posits the right microstructure, since the EFEs are likely to be recoverable in many theories of quantum gravity that feature a Lorentzian manifold and a microstructure that admits thermodynamic equilibration.

\subsection{Entanglement gravity}\label{Entanglement gravity}
Already in his thermodynamic gravity, Jacobson alludes to the possibility that the entropy associated with the Rindler horizon might originate in entanglement.
\footnote{It is disputed whether entanglement entropy is genuine thermodynamic entropy and the approaches to the EFEs based on entanglement entropy might as a consequence not be examples of \textit{thermodynamic} gravity. There are indications that entanglement entropy is indeed connected to thermodynamic entropy \citep{kaufman_quantum_2016} and regardless, the entanglement approaches to EFEs considered below at most assume an entanglement thermodynamics -- genuine or not -- analogous to ordinary thermodynamics (see \citet{alishahiha_entanglement_2013} for more details on entanglement thermodynamics).} Entanglement entropy measures how entangled the degrees of freedom within some subsystem are with the degrees of freedom outside it.\footnote{Formally, the entanglement entropy, $S_B$ of a quantum subsystem living in the manifold subregion $B$ is defined as $S_B = - \rho_B \log(\rho_B)$. Here $\rho_B$ is the reduced density matrix for the quantum subsystem.} In the following, these subsystems will be associated with subregions of the manifold on which the QFT is defined. The first approach to entanglement gravity to be considered below assumes the AdS/CFT correspondence, whereas the second approach considered -- also due to Jacobson -- trades the assumption of AdS/CFT with an equilibrium assumption and some lessons from many-body systems.

The AdS/CFT correspondence is a conjectured duality between theories with gravity in asymptotically Anti-de Sitter spacetime (AdS)\footnote{Anti-de Sitter spacetimes are vacuum solutions to the EFEs with a negative cosmological constant, i.e. solutions whose curvature is the other way around as compared to the current best cosmological models. It appears, therefore, that the AdS/CFT correspondence does not obtain in the actual world.} and conformal field theories (CFT) defined on non-dynamical Lorentzian manifold \citep{maldacena_large-n_1999}, i.e. the matter fields of the CFT do not couple to the metric of the manifold and the metric does as a consequence not change.
\footnote{The correspondence has its origin in string theory and conjectures more precisely that certain closed string theories in asymptotically AdS are dual to certain CFTs defined on a fixed background identical to the asymptotic boundary of the dual AdS spacetime. See \citet{butterfield_conceptual_2016} for a philosophical introduction to the AdS/CFT correspondence.} Being a duality, the two theories are conjectured to be empirically equivalent, but different ways to represent the same physics.\footnote{See for instance \citet{de_haro_dualities_2017} for more details about the notion of duality in the context of the AdS/CFT correspondence. All metaphysical questions, for instance whether the AdS side and the CFT side represent metaphysically distinct possible worlds or not will be set aside since both worlds---by definition of the notion of duality---would be empirically indistinguishable (see \citet{read_interpretation_2016} and \citet{le_bihan_duality_2018} for an overview of these metaphysical questions).} In CFTs, a Clausius-like relation can be derived as a purely formal result: For any ball shaped region of a CFT, the change in entanglement entropy of a region is proportional to its change in hyperbolic energy \citep{casini_towards_2011}.\footnote{Formally we have $dS_B = \delta E^{hyp}_B$ if
\[
\delta E^{hyp}_B = \pi \int_B d^{d-1} x \frac{R^2-(\vec{x}-\vec{x_0})^2}{R} \delta \expval{T^{tt}(x)}
\]
where $R$ is the radius of the ball shaped region and $\delta \expval{T^{tt}(x)}$ is the variation of the energy density.} Since this is a formal result of CFTs, the relation holds always and not only for equilibrium states. In the limit where the AdS side can be approximated by semi-classical gravity, also a relation between entropy and areas can be shown to obtain. More precisely, this so-called Ryu-Takayanagi formula states that areas on the AdS side are proportional to the entanglement entropy of subsystems on the CFT side \citep{ryu_holographic_2006}.\footnote{In mathematical terms, the Ryu-Takayanagi formula takes the form: $S_B = \text{Area}( \tilde{B})/4G$ where $S_B$ is the entanglement entropy of a subregion of the manifold, $B$, on the CFT side and $\tilde{B}$ is the area on the AdS side related to the region $B$ on the CFT side. For further details about the Ruy-Takayanagi formula including the evidence for it within the AdS/CFT correspondence, see \citet{rangamani_holographic_2017}.} Using the Ryu-Takayanagi formula (and nothing else of the AdS/CFT correspondence), one can translate the CFT relation between entanglement entropy and hyperbolic energy to the AdS side where it proves to be equivalent to the linearized vacuum EFEs \citep{lashkari_gravitational_2014,faulkner_gravitation_2014}. 
Taking into account corrections to the Ryu-Takayanagi formula due to entanglement on the AdS side \citep{lewkowycz_generalized_2013}, the relation between entanglement entropy and hyperbolic energy instead translates into linearized EFEs with matter content and has indications of a generalization to the full EFEs \citep{swingle_universality_2014}.

While still in need of further development, the argument suggests a way to conceive of the EFEs as entanglement dynamics and this sounds promising for the present purposes. Any quantum degrees of freedom can be entangled. Entanglement can, in other words, be realized in any quantum system with two or more degrees of freedom. Arguably, entanglement will, as a consequence, feature in any theory of \textit{quantum} gravity regardless the details of its proposed microstructure; it just has to be quantum. Obviously, a number of assumptions go into the relation between the EFEs and entanglement, but if these can be shown to be sufficiently generic, then general relativity -- by its assumed relation to the EFEs -- might inherit the generic nature of entanglement in theories of quantum gravity. However, care must be taken when it comes to dualities. 
The result so far is that if the Ryu-Takayanagi formula with bulk entanglement corrections holds, then the EFEs (on the AdS side) can instead be encoded as a constraint on the relation between hyperbolic energy and entanglement entropy (on the CFT side). 
Several authors have argued that the AdS/CFT correspondence, as an exact duality, cannot testify to the \textit{emergence} of gravity (e.g.  \citet{teh_holography_2013}, \citet{rickles_ads/cft_2013}, \citet{dieks_emergence_2015}, and \citet{butterfield_conceptual_2016}). The AdS side and CFT side are simply different ways to represent the same physics: ``the presence of a duality precludes the phenomenon of emergence" \citep[110]{de_haro_dualities_2017}. Just as we may conceive of the EFEs as entanglement dynamics in such circumstances, so it is possible to conceive of entanglement dynamics in terms of the EFEs. The situation, therefore, is different from induced and thermodynamic gravity. These showed how the EFEs can be derived as an effective description holding in a particular limit -- thermodynamic or first loop order -- even if gravity is absent at the level of description of quantum gravity and thus only manifests itself as a large scale phenomenon. In these accounts, the EFEs are in this sense emergent. Such emergence is precluded for exact dualities and thus when the Ryu-Takayanagi formula relates areas of surfaces in one theory (the AdS side) to entanglement entropies in another theory (the CFT side). The Ryu-Takayanagi formula simply takes the form of a relation between representations like the entry of a dictionary. However, the concern here is not the emergence of gravity, but rather indications that general relativity is multiply realizable. If the EFEs are equivalent to some constraint on entanglement entropy and this constraint is multiply realizable, then this characteristic must, due to the duality, be shared  by the EFEs and therefore general relativity: What is multiply realizable on one side cannot be a physical kind on the other.

As already argued, entanglement is to be expected in any theory of \textit{quantum} gravity. However, the additional assumptions that go into entanglement gravity -- the Ryu-Takayanagi formula and the relation between entanglement entropy and hyperbolic energy -- are not as innocent. Assuming the full battery of the AdS/CFT correspondence, the relation between entanglement entropy and hyperbolic energy is a formal fact -- it holds for all CFTs -- and the Ryu-Takayanagi formula is proven to hold in the limit where the AdS side can be approximated by semi-classical gravity \citep{casini_towards_2011,lewkowycz_generalized_2013}. The problem with this line of argument, however, is the AdS/CFT correspondence itself. The correspondence was originally conceived within string theory and is therefore deeply rooted in this particular approach to quantum gravity. In order for the EFEs to potentially inherit the versatility of entanglement, their relation cannot rely on idiosyncratic aspects of string theory. Thus, one must depart from string theory, and therefore the framework of the AdS/CFT correspondence, and seek more generic reasons for the Ryu-Takayanagi formula\footnote{There are indications, though conjectural, that the Ryu-Takayanagi formula generalizes to other prospective theories of quantum gravity: causal set theory \citep{sorkin_entanglement_2018}, loop quantum gravity \citep{smolin_holographic_2016,han_loop_2017}, and the more general framework of generic group field theories employing a relation to tensor networks with promises of even further generalization \citep{chirco_group_2018}. But the worries raised in section \ref{Towards an argument} might be raised anew, if too much emphasis is put on our current field of prospective quantum gravity theories.} and the relation between entanglement entropy and hyperbolic energy to hold. 

One such attempt to establish a relation between the EFEs and entanglement inspired by 
the discoveries of AdS/CFT\footnote{Jacobsen explicitly refers to \citet{lashkari_gravitational_2014} and \citet{faulkner_gravitation_2014} as central sources of inspiration (2016, 1).} but without stipulating the AdS/CFT correspondence
is proposed by \citet{jacobson_entanglement_2016}.\footnote{For the sake of brevity, we shall here limit the attention to \citet{jacobson_entanglement_2016}, but there are several other approaches to the EFEs from entanglement that are inspired by AdS/CFT without stipulating it, e.g. \citet{verlinde_emergent_2017} and \citet{cao_bulk_2018}.} Jacobson also considers a ball shaped region of a Lorentzian manifold, but now with a generic QFT living on it. Without the AdS/CFT correspondence, the Ruy-Takayanagi formula is reinterpreted as a leading order proportionality between the area and entanglement entropy of the ball. Jacobson motivates this proportionality at leading order with the observation that the primary contribution to the entanglement entropy comes from short range entanglement over the boundary of the ball which accounts for the area scaling.\footnote{For the purposes of Jacobson's argument, it is even sufficient that the UV part of the entanglement entropy is proportional with area, an assumption that looks even less contestable.} The relation between entanglement entropy and hyperbolic energy is motivated by the assumption that the QFT under consideration has a UV fixed point such that it is conformally invariant at sufficiently small distances and that the ball is small enough to be conformally invariant throughout. Finally, due to the details of the argument, Jacobson must assume that the ball is at an entanglement equilibrium such that the first order variation of the entanglement entropy vanishes.
\footnote{Jacobson also uses the (contracted) Bianchi identity which assumes that the connection is metric compatible. Though this is a very common assumption in general relativity, reality might prove to be otherwise, but we will not pursue this further here. Thank you to one of the anonymous reviewers of \textit{Synthese} for raising this issue.} 
However, given these elements, the full EFEs can be derived. Thus, general relativity can be recovered as the result of entanglement equilibrium under the assumption of a UV fixed point and leading order area scaling of the entanglement entropy.

Being freed from the AdS/CFT correspondence and therefore string theory, Jacobson's derivation suggests a way for general relativity to inherit the generic nature of entanglement. Jacobson's assumed UV fixed point and entailed conformal invariance can be motivated as a necessary condition for any fundamental theory. This follows since the lack of a UV fixed point -- often as connected with non-renormalizability -- is usually taken as evidence that the theory is merely an effective theory that fails to incorporate some underlying microstructure.\footnote{\citet{crowther_renormalizability_2017} argue that a theory of quantum gravity does not have to be fundamental and that UV-completeness cannot be assumed for quantum gravity. We will disregard this subtlety here.} As such, Jacobson's assumed UV fixed point should exist for any theory of quantum gravity that aims to be a fundamental theory. Also leading order area scaling of entanglement entropy is a common result, and seems for instance to be generic in quantum many body systems \citep{eisert_colloquium:_2010}. 
In addition, such area scaling suggests that the \citet{bekenstein_universal_1981} bound on entropy is saturated to leading order, something that accords well with the view that ``holographic principle is generally viewed as an important guiding principle in the search for a theory of quantum gravity" \citep{linnemann_hints_2018}.
\footnote{Some care must be taken when it comes to the holographic principle since its derivation assumes parts of general relativity. Using it to argue that general relativity is multiply realizable might therefore be circular. This, however, it not the place to discuss this issue further. See \citep[section ii]{bousso_holographic_2002} for a discussion of the various routes to the holographic principle.} Finally, the result that a small ball -- a small local system -- should be at an entanglement equilibrium after some time is supported by quantum thermalization in generic many-body systems \citep{kaufman_quantum_2016}. Here, it is found that small local systems over time thermalize, i.e. enter an equilibrium with respect to entanglement entropy even when the full system remains in a pure state. Both of these arguments from quantum many-body physics rely on macroscopic behaviour that -- in analogy to the case of thermodynamics -- promises to be sufficiently insensitive to the details of the underlying microstructure; their nature is unimportant for the mentioned effects.

In summary, Jacobson's derivation suggests how the EFEs can obtain as an effective description due to generic entanglement dynamics satisfying the entanglement equilibrium and leading order area scaling of entanglement. With these elements of the derivation and entanglement itself promising to be generic features in quantum gravity, this again entails that general relativity can be multiply realizable in quantum gravity. The recovery of general relativity in quantum gravity theories is unsurprising due to this relation to entanglement that in turn should feature in any theory of \textit{quantum} gravity.

\section{Conclusion}\label{Conclusion}
Does such a recovery of general relativity in a theory of quantum gravity ensure that the microstrcuture it posits finds a recognizable counterpart in the true theory of quantum gravity? Based on a functionalist perspective, this paper has defended a negative answer to the question: General relativity can be recovered from elements that are multiply realizable or at least generic in nature in quantum gravity, and one can therefore expect that general relativity can be recovered in theories of quantum gravity that describe other possible worlds than the actual one. A theory of quantum gravity can recover general relativity, while being completely wrong about the microstructure it posits.

The argument was founded on three research schemes -- induced gravity, thermodynamic gravity, and entanglement gravity -- where the EFEs are derived from generic elements that do not invoke any particulars of the type of microstructure realizing the EFEs. With the insensitivity to the microstructure, these accounts suggest how the EFEs can be multiply realizable. With reference to string theory, it was indicated what is meant, when a theory of quantum gravity is said to recover general relativity, and the three accounts each provide an indication of how general relativity can be multiply realizable in quantum gravity. Given the character of each case study, I conjecture that we can expect many prospective theories of quantum to recover general relativity even though the microstructure they posit are not that of the actual world.

This impedes to a degree the role of general relativity in non-empirical confirmation of quantum gravity. If general relativity, as argued here, is multiply realizable in quantum gravity, then we should not be too impressed by a theory of quantum gravity where general relativity can be recovered in some limit of the theory. The recovery of general relativity cannot serve as the ultimate arbiter that decides which theory of quantum gravity that is worthy of pursuit, even though it is of course not irrelevant either \textit{qua} quantum \textit{gravity}. Returning where we started, it is arguably essential that string theory recovers general general relativity for its prospects as a theory of quantum gravity, but it does not establish that string theory must have some truth to it.

\subsection*{\small Acknowledgements}
\footnotesize I would like to express my gratitude to Niels Linnemann, Kian Salimkhani, Astrid Rasch, Richard Dawid, Sorin Bangu, and two anonymous reviewers of \textit{Synthese} for valuable feedback on and helpful discussion of earlier drafts of this paper. I also send my thanks for constructing comments to the participants at the Spacetime Functionalism Workshop (University of Geneva) and 1st Scandinavian Workshop on the Metaphysics of Science (NTNU) where earlier versions of this paper was presented.

\bibliography{bib}
\end{document}